\documentclass[11pt]{article}
\usepackage{color}
\usepackage{amsfonts}
\usepackage{amssymb}
\usepackage{amsthm}
\usepackage{amsmath}
\usepackage{graphics}
\usepackage{graphicx}
\usepackage{a4wide}
\usepackage{url}

\newcommand{\R}{{\mathbb R}}

\newcommand{\Z}{\mathbb{Z}}
\newcommand{\Ceq}{\stackrel{+}{=}}
\newcommand{\cS}{{\cal S}}

\begin{document}

\theoremstyle{definition}
\newtheorem{Q}{Question}

\newtheorem{Con}{Condition}

\theoremstyle{remark}
\newtheorem{Rem}{Remark}

\theoremstyle{plain}
\newtheorem{Def}{Definition}
\newtheorem{Lem}{Lemma}
\newtheorem{Prop}{Proposition}
\newtheorem{Thm}{Theorem}
\newtheorem{Cor}{Corollary}
\newtheorem{Post}{Postulate}

\newcommand{\peq}{\stackrel{+}{=}}
\newcommand{\pleq}{\stackrel{+}{\leq}}
\newcommand{\pgeq}{\stackrel{+}{\geq}}
\newcommand{\mbbZ}{\mathbb{Z}}
\newcommand{\Perp}{\perp \! \! \! \perp}

\title{Justifying additive-noise-model based 
causal discovery via algorithmic information theory}

\author{Dominik Janzing and Bastian Steudel}

\author{
Dominik~Janzing$^1$ and Bastian Steudel$^2$\\
${}$\\
{\small  1) Max  Planck Institute for Biological Cybernetics} \\
{\small T\"ubingen, Germany} \\
${}$\\
{\small 2) Max  Planck Institute for Mathematics in the Sciences}\\
{\small Leipzig, Germany}
}

\date{October 09, 2009}

\maketitle

\abstract{A recent method for causal discovery  is in many  cases able to
infer whether $X$ causes $Y$ or $Y$ causes $X$ for just two observed variables $X$ and $Y$.
It is based on the observation that there exist (non-Gaussian) joint distributions $P(X,Y)$ for which
$Y$ may be written as a function of $X$ up to an additive noise term
that is independent of $X$  and no  such model exists from $Y$ to $X$.
Whenever this is the case, one prefers the causal model $X\rightarrow Y$. 

Here we justify this method by showing  that 
the causal hypothesis $Y\rightarrow X$ is unlikely 
because 
it requires a specific tuning  between $P(Y)$ and $P(X|Y)$ to generate a  distribution that admits an
additive noise model from $X$ to $Y$. To quantify the amount of tuning required we derive lower bounds 
on the {\it  algorithmic} information shared by $P(Y)$  and  $P(X|Y)$.
This way, our justification is consistent with recent approaches for using algorithmic information theory 
for causal reasoning. 
We extend this principle to the case where $P(X,Y)$ {\it almost} admits  an additive noise model.

Our results suggest that  the above conclusion is more reliable if the complexity of $P(Y)$
is high. 
}

\section{Additive noise models in causal discovery}

Causal inference from  statistical data is a field of research that obtained 
increasing interest in recent years. To infer causal  relations among several random variables
by purely observing their joint distribution is unsolvable from the point  of view of traditional statistics.
During the 90s, however, it was more and more believed that also
non-experimental data contain at least {\it hints} on the causal directions. 
The most important postulate that links the observed statistical dependencies on the one hand 
to the causal structure
(which is here assumed to be a DAG, i.e., a directed  acyclic graph) on the  other hand
is the causal Markov condition \cite{Pearl:00}.
It states that every variable is conditionally independent of its  non-effects, given its causes. 
If  the joint distribution $P(X_1,\dots,X_n)$ has a  density $p(x_1,\dots,x_n)$
with respect to some product measure, then  the density
factorizes  \cite{Lauritzen}
into
\[
p(x_1,\dots,x_n)=\prod_{j=1}^n p(x_j|pa_j)\,,
\]
where $p(x_j|pa_j)$  denotes the conditional probability density of $X_j$,  given the values $pa_j$ of its parents $PA_j$.

The Markov condition already rules out some DAGs as being incompatible with the 
observed conditional dependencies. However, usually a large set of DAGs still is compatible. 
In particular, for $n$ variables, 
there are $n!$ DAGs that are consistent 
with every joint distribution because they do not impose any conditional independence.
They are given by defining an  order $X_1,\dots,X_n$ and drawing an error from $X_i\rightarrow X_j$ for every
$i<j$. For this reason, additional  inference rules  are required to choose the most  plausible ones among the
compatible DAGs. Spirtes at  al.~\cite{Spirtes1993} and Pearl \cite{Pearl:00} use the causal faithfulness principle
that prefers those DAGs for which the causal Markov condition imposes all the observed {\it in}dependencies.  In other words, 
it is considered unlikely that independencies are due to particular (non-generic) choices of the conditionals
$p(x_j|pa_j)$.  The  underlying idea is, so to speak, that ``nature chooses'' the conditionals independently  from  each other,
while  the generation of additional independencies (that are not imposed by the structure of the DAG) 
would require to mutually adjust these conditionals. A more general perspective on
such an independence 
assumption has been provided by
 Lemeire and Dirkx \cite{LemeireD} who stated the following principle:

\begin{Post}[Algorithmic independence of conditionals]${}$\\ \label{IC}
If the true causal structure is given by the directed acyclic graph $G$ with random variables 
$X_1,\dots,X_n$ as nodes, 
the shortest description of the joint density $p(x_1,\dots,x_n)$ 
is given by separate  descriptions  of 
the conditionals\footnote{For sake of simple terminology, we also consider the density $p(x_j)$ of  parentless nodes as a ``conditional'', given an empty set of variables.} $p(x_j|pa_j)$. 
\end{Post}
In \cite{LemeireD} the description length has been defined in  terms of algorithmic information, also  called ``Kolmogorov complexity'' (the details will be explained in Section~\ref{AlIn}). There the postulate is mainly used to justify the causal faithfulness assumption \cite{Spirtes1993}, since it rules out mutual adjustments among conditionals like those required for unfaithful distributions. 
However, in \cite{Algorithmic} it has been argued that the complete determination of the joint distribution is never feasible which makes it hard  to give empirical content to it. Moreover,  \cite{Algorithmic} shows that Lemeire and Dirkx's principle can be seen  as an implication  of a general framework for causal inference via algorithmic   information.
There, the postulate is rephrased in a way that avoids the complexity of conditionals  and uses only empirical observations. Furthermore, the general framework imposes many causal inference rules yet to be discovered. Here we focus on a method \cite{Hoyer} that yielded quite encouraging results on real data sets and show that it also can be justified via algorithmic information theory. We briefly rephrase the idea of \cite{Hoyer} for the special case of two real-valued variables $X$ and $Y$. To this end we introduce the following terminology:
\begin{Def}[Additive noise model]${}$\\
The joint  density $p(x,y)$  of two real-valued random variables $X$ and $Y$ 
is said to admit an  additive noise model
from $X$ to $Y$ if there is a measurable function $f:\R\rightarrow \R$ such that
\begin{equation}\label{NLiModel}
Y=f(X)+E\,,
\end{equation}
where $E$ is some unobserved noise variable that is statistically independent
of $X$. 
The joint density thus is of the form
\[
p(x,y)=p_X(x)p_{E}(y-f(x))\,,
\]
where $p_X(x)$ is the density of $X$ and $p_{E}(e)$ the density of $E$. 
\end{Def}

Whenever this causes no confusion, we will drop the indices and write 
$p(x)$ instead of $p_X(x)$ and, similarly, write $p(y-f(x))$. We will write $p_X$ if we want to emphasize that
we refer to the entire  density and not one specific value $p(x)$.

It can be shown  \cite{Hoyer} that for generic choices of $f$, distribution of the noise, and distribution of $X$, there is no additive noise model from $Y$ to $X$.
In other  words, if causality in nature would 
always be of the form of additive noise models (which is certainly
not the case\footnote{For instance, \cite{Zhang_UAI} discusses an interesting generalization.}),
we could almost always identify causal directions because a joint distribution that admits an 
additive  noise model in the true direction usually does not admit one in  the wrong direction.
This paper addresses the question whether a causal  structure $Y\rightarrow X$ that is {\it not} of  the form
of an additive noise model could induce a joint distribution that admits an additive noise model in the  wrong direction
(i.e., from $X$  to $Y$).
The basic observation of this paper is  that this would be a rare coincidence because it requires
that $p_Y$  (which would be the distribution of the cause) and the transition probabilities $p_{X|Y}$ 
(which describes the effect generating the relation between cause and effect) satisfy an untypical relation
that makes this scenario unlikely.
However, instead  of deriving probability values for such  a coincidence (which required to assign  priors on probability  distributions) we will take a non-Bayesian view and follow the algorithmic information theory approach developed in \cite{Algorithmic}  and \cite{LemeireD}. 
The following lemma makes explicit what kind of coincidence is meant:

\begin{Lem}[Relation between $p_Y$ and  $p_{X|Y}$]${}$\\ 
Let  $p(x,y)$ be positive definite and let $f$ as well as all logarithms  of marginal and conditional densities
be two times  differentiable. If $p(x,y)$ admits an additive noise model from $X$ to $Y$, 
then the marginal $p(y)$ and the conditional $p(x|y)$ are related via the differential equation
\begin{equation}\label{DGL}
\frac{\partial^2}{\partial y^2} \log p(y) =- \frac{\partial^2}{\partial y^2} \log p(x|y) -\frac{1}{f'(x)} \frac{\partial^2}{\partial x \partial y} \log p(x|y)\,.
\end{equation}
\end{Lem}

\vspace{0.3cm}
\noindent
Hence we have 
\[
\log  p(y) = -\int_0^y \int_0^{y''} \frac{\partial^2}{\partial y^2} \log p(x|y') -  \frac{1}{f'(x)} \frac{\partial^2}{\partial x \partial y}\log p(x|y') dy' dy'' + ay+b\,,
\] 
where $b$ is determined by $\int p(y) dy=1$. Since the equation has to be valid for all $x$, we can choose
an arbitrary $x_0$  with $f'(x_0)\neq 0$. 
Then $p_Y$  can already be determined from $f'(0)$, the function $y\mapsto p(x_0|y)$  and $a$.
Given the conditional  $p_{X|Y}$, the tupel $(x_0,f'(x_0))$ and $a$ 
are sufficient to describe 
the marginal $p_Y$. In general, these are much fewer parameters  than those required 
for describing $p_Y$  without  knowing $p_{X|Y}$. This
already suggests that $p_Y$ and  $p_{X|Y}$ have  algorithmic information in common because 
knowing $p_{X|Y}$ shortens the description of $p_Y$.

However, assume we know that  $p_{XY}$ belongs to the family of bivariate Gaussians. Then it admits  an  additive noise model in both directions and both causal directions are possible. This is consistent with  the fact that our argument above fails in  this case because $a$ and $f'(x_0)$ then coincides with the information that also would be required to describe $p_Y$ {\it without} knowing $p_{X|Y}$. 
To see this, 
set
\[
\log p(x)\stackrel{+}{=} \frac{(x-\mu_X)^2}{2\sigma_X^2}\,,
\]
where $\stackrel{+}{=}$ denotes equality up to a term that neither depends on $x$ nor on $y$.
Furthermore, let
\[
\log p(y|x)\stackrel{+}{=}\frac{(y-cx-\mu_E)^2}{2\sigma_E^2}\,,
\]  
with the notation $c:=f'(x_0)$.
We then  get
\[
\log p(y)\stackrel{+}{=}\frac{(y-\mu_X-\mu_E)^2}{2(c^2 \sigma_X^2+\sigma_E^2)}.
\]
Hence,
\[
\log p(x|y)\stackrel{+}{=} \frac{(x-\mu)^2}{2\sigma_X^2}+\frac{(y-cx)^2}{2\sigma_E^2} -\frac{(y-\mu_X-\mu_E)^2}{2(c^2\sigma_X^2+\sigma_E^2)}\,,
\]
which implies
\[
\frac{\partial^2}{\partial x \partial y} \log p(x|y) \stackrel{+}{=}  -\frac{c}{\sigma_E^2} =:\alpha\,,
\]
and
\[
\frac{\partial^2}{\partial y^2} \log p(x|y) \stackrel{+}{=}  -\frac{1}{\sigma_E^2} =:\beta\,.
\]
The constants $\alpha$ and $\beta$ can be derived from observing $p(x|y)$, but to determine the second derivative of
$\log p_Y$ one needs to know $c$  since eq.~(\ref{DGL}) imposes
\begin{equation}\label{Gauss}
\frac{\partial^2}{\partial y^2} \log p(y) = \beta - \frac{1}{c}\alpha\,.  
\end{equation}
To determine $p_Y$ completely, we also need to know the first derivative
\[
a:=\frac{\partial}{\partial y}\log p(y=0)=-\frac{\mu_Y}{\sigma_Y^2}\,,
\]
if $\mu_Y$ denotes the mean of $Y$.
Moreover, we  observe that $c$ specifies the standard deviation  $\sigma_Y$ of $Y$ because
the  left hand side of eq.~(\ref{Gauss}) is given by $-1/\sigma_Y^2$.
This shows, that given $p_{X|Y}$, we still need to describe the two parameters $\mu_Y$ and 
$\sigma_Y$. These are exactly  the two parameters that describe the Gaussian  $p_Y$  
also {\it without} knowing $p_{X|Y}$. Hence,
knowing $p_{X|Y}$ is worthless for the description of
$p_Y$. 

The intuitive arguments above show that 
knowing  $p_{X|Y}$ makes the description of $p_{Y}$ shorter 
except for some rare cases where $p_Y$ already  has a  short description.
Formal statements of this kind, however, require the specification of the accuracy up to which $p_Y$ and $a$ are described.  

The paper is structured as follows.  In Section~\ref{AlIn} we briefly rephrase algorithmic information theory based
causal inference as developed in  \cite{Algorithmic}. In  Section~\ref{ANC} we show that additive noise models
from $X$ to $Y$ induce densities $p_Y$ and  $p_{X|Y}$  that have algorithmic information in common.  
In Section~\ref{Dis} we consider additive noise models over finite fields and show  that $p_Y$ and $p_{X|Y}$ also share algorithmic information if the  distribution is only {\it close 
to} an additive noise model from $X$ to $Y$. Since our bounds on the information shared by these objects
depend on the Kolmogorov complexity of $p_Y$ (which cannot be determined) we 
discuss a method to estimate the latter in
Section~\ref{EstK}. Section~\ref{secEmp} and Section~\ref{Con} discuss  how to apply the insights gained from the discrete case to empirical and to continuous distributions respectively.

\section{Algorithmic information theory and the causal principle} 

\label{AlIn}

Reichenbach's Principle of Common Cause \cite{Reichenbach} is meanwhile the cornerstone of 
causal reasoning  from statistical data:
Every statistical dependence between two random variables $X$ and $Y$ 
indicates at least  one of the three causal relations (1) ``$X$ causes $Y$'', (2) ``$Y$ causes $X$'', 
or (3) is a common cause $Z$  influencing both $X$ and $Y$.
As an extension of this principle, we have  argued  \cite{Algorithmic} that 
causal inference is not always based on {\it statistical} dependencies.
Instead, similarities between single objects also indicate causal links (e.g., if two T-shirts produced by different companies have the same
sophisticated pattern we would not believe that the designer came up with the patterns independently).
We have therefore postulated the ``causal principle'' stating that there is a causal link between two objects 
whenever the joint description of them is shorter than the concatenation of their separate descriptions.

To formalize this, we first introduce some  concepts of algorithmic information theory \cite{Vitanyi97}.
Let $s,t$ be two binary strings that describe the observed objects and let $K(s)$ denote the
algorithmic information (or ``Kolmogorov complexity''), i.e., the length of the shortest program that generates $s$ 
on a universal Turing machine  \cite{KolmoOr,Solomonoff,ChaitinF,Chaitin}. 
Let  $K(s|t)$ denote the length of the shortest program that generates $s$ from  the input $t$.
Then we define
\cite{GacsTromp}:

\begin{Def}[algorithmic mutual information]${}$\\
Let $s,t$ be two binary  strings. 
Then the algorithmic mutual information between $s$ and $t$ reads
\begin{equation}\label{AlgM}
I(s:t):=K(t)-K(t|s^*)\stackrel{+}{=}K(s)+K(t)-K(s,t)\,,
\end{equation}
where $s^*$ denotes the shortest program that computes $s$ and $K(s,t)$ is  the length of the shortest program generating 
the concatenation of $s$ and $t$. 
\end{Def}

As usual in algorithmic information theory, all (in)equalities are only
understood up to a constant that depends on the Turing machine \cite{Vitanyi97}. For this reason, we write
$\Ceq$ instead of $=$. Since  $s$ can be  computed  from $s^*$, but  usually not  vice versa,
we have 
\begin{equation}\label{star}
K(t|s^*)\stackrel{+}{\leq} K(t|s)\,.
\end{equation}
We will later also need the conditional version of (\ref{AlgM}), see
\cite{GacsTromp}:

\begin{Def}[conditional algorithmic mutual information]${}$\\  
Let $s,t,v$ be  binary strings. Then  the  conditional algorithmic mutual information
reads
\begin{equation}\label{cMI}
I(s:t|v):=K(t|v)-K(t|s,K(s|v),v)\Ceq K(s|v)+K(t|v)-K(s,t|v)\,.
\end{equation}
\end{Def}
Eq.~(\ref{AlgM}) is formally similar to the statistical mutual information 
\[
I(X:Y):=H(Y)-H(Y|X)=H(X)+H(Y)-H(X,Y)\,,
\]
phrased in terms of the Shannon entropy $H(\cdot)$.
Reichenbach's principle can then be rephrased as: 
\begin{quote}
``$I(X:Y)> 0$ indicates  that there is at least
one of the  three possible causal links between $X$ and  $Y$.''
\end{quote}
In analogy to this principle, we have postulated in \cite{Algorithmic}:

\begin{Post}[Causal Principle]${}$\\ \label{CP}
Let $s$ and $t$ be binary strings that formalize  the descriptions of two objects in  nature.
Whenever 
\[
I(s:t)\gg 0\,,
\]
there is a causal link between the two objects $s$ and $t$ in the sense that
$s\rightarrow t$ or $t\rightarrow s$ or there is a third object $u$ with
$s\leftarrow  u \rightarrow t$.
\end{Post}

Here,   it   is up  to the researcher's decision how to set the threshold above which a dependence is considered
significant.
This is similar  to setting the significance value in a statistical test.
 
Note that the condition
$
K(t)-K(t|s) \gg 0
$ implies $I(s:t)\gg 0$ due to ineq.~(\ref{star}). 
We will work with the former  condition since it is easier to test. 

\begin{figure}
\centerline{\includegraphics[scale=0.20]{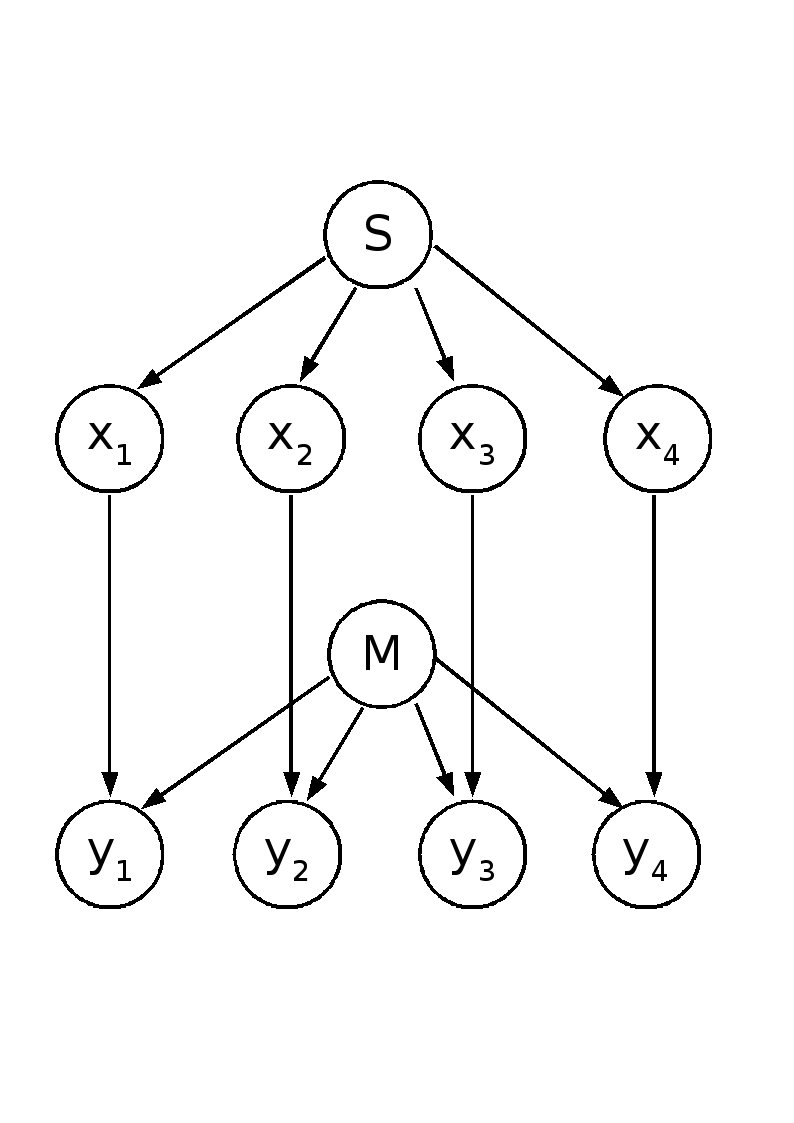}}
\caption{{\small Causal structure obtained by resolving the causal structure $X\rightarrow Y$ 
between the random variables $X$ and $Y$ 
into causal relations among single events}\label{SM}}
\end{figure}

To interpret Postulate~\ref{IC} as a special case of Postulate~\ref{CP},  
we consider the following model \cite{Algorithmic} of a causal structure $X\rightarrow Y$ for two random variables
$X$ and $Y$. Take as the two objects in nature a source $S$ that generates $x$-values according to $p(x)$  and a machine $M$ that takes $x$-values as input and generates $y$-values according to $p(y|x)$ (see Figure~\ref{SM}).

If $S$ and $M$ have been designed independently, their optimal joint description should be given by separate descriptions of $S$ and  $M$. However, the only feature of $S$ that is relevant for our observations
is given by the distribution  of $x$-values, i.e., $p_X$. Similarly, $p_{Y|X}$  is the only relevant feature
of $M$. These features are directly  given by observing the $x$ and the  $y$-values after infinite sampling.
We therefore consider the algorithmic dependencies between $p_X$ and  $p_{Y|X}$.
Since  the objects of our descriptions will be probability distributions, we  introduce the following concept:

\begin{Def}[computable functions and distributions]${}$\\
Let  $\cS$ denote some  subset of $\R^k$.
A function $f:\cS\rightarrow \R$ is computable if there is a program that computes $f(x)$ up to a precision $\epsilon>0$ for every input $(x,\epsilon)$, for which $x$ has a finite description. Then  $K(f)$ denotes the length of the
shortest program of  this kind. A probability distribution  on a
finite probability space $\cS$ is called computable if its density is a computable function.
\end{Def}
In the following section we apply the concepts introduced above to the case of strictly positive continuous 
densities $p(x,y)$.

\section{Algorithmic dependencies induced by additive  noise models}

\label{ANC}

We  have already argued  that an additive noise model from $X$ to $Y$ makes the causal structure 
$Y\rightarrow X$ unlikely because $p_Y$ and $p_{X|Y}$  then satisfy the non-generic relation
of eq.~(\ref{DGL}). We now express this fact in terms of algorithmic information theory:

\begin{Thm}[algorithmic dependence induced by an additive noise model]${}$\\   \label{CAI}
Let $p(x,y)$ be a two-times differentiable computable strictly positive probability density over $\R^2$. 
If $p(x,y)$ 
admits an additive noise model from
$X$ to $Y$ with a computable differentiable function $f$, then 
\[
I(p_Y: p_{X|Y}) \stackrel{+}{\geq} K(p_Y)- K(y_0,\psi'(y_0))-K(x_0,f'(x_0)) \,
\]
where $x_0$ and $y_0$ are  arbitrary computable $x$- and $y$-values, respectively and $\psi (y):=\log p(y)$.
\end{Thm}

\noindent
Proof: Eq.~(\ref{DGL}) expresses the second derivative $( \log p_Y)''$ in terms of $p_{X|Y}$ and $f'(x_0)$.
Hence,
\begin{equation}\label{condDer}
K((\log p_Y)''|p_{X|Y}) \stackrel{+}{\leq } K(x_0,f'(x_0))\,.
\end{equation}
We have by definition
\begin{equation}\label{muY}
I(p_Y:p_{X|Y})\stackrel{+}{=}  K(p_Y)-K(p_Y|p^*_{X|Y})\stackrel{+}{\geq } K(p_Y)-K(p_Y|p_{X|Y})\,.
\end{equation}
The density $p_Y$ is already determined by  $(\log p_Y)''$ and the first derivative $\psi' (y_0)$ for some $y_0$ because $\log p_Y(y_0)$ then follows from normalization. Therefore,
\[
K(p_Y|z)\stackrel{+}{=} K((\log  p_Y)''|z)+K(\psi (y_0)|z)\,,
\] 
where $z$ is some arbitrary prior information. 
Using $z=p_{X|Y}$, the right  hand term  of ineq.~(\ref{muY}) yields
\begin{eqnarray*}
I(p_Y:p_{X|Y}) &\stackrel{+}{\geq} & K(p_Y)-   K( (\log p_Y)''|p_{X|Y})-K(y_0,\psi'(y_0)|p_{X|Y}) 
\\
&\stackrel{+}{\geq } &
K(p_Y)-K(x_0,f'(x_0)|p_{X|Y})-
 K(y_0,\psi '(y_0)|p_{X|Y})  \\&\stackrel{+}{\geq }&
K(p_Y)-K(x_0,f'(x_0))
- K(y_0,\psi '(y_0)) 
\,,
\end{eqnarray*}
where  the second inequality is due to ineq.~(\ref{condDer}).
$\Box$

\vspace{0.3cm}
The interpretation of Theorem~\ref{CAI} raises two problems: First, we cannot determine the exact ``true''  probabilities\footnote{It is, anyway, a philosophical problem to what extent they are well-defined.} 
from  the observations, and second, we do not expect these probabilities  to be computable, and
hence it 
required an infinite amount of 
information to describe $p_Y$ and $p_{X|Y}$ if
we could. 
As already pointed out in \cite{Algorithmic}, algorithmic dependencies
among the {\it  empirical} distributions $q_Y$ and $q_{X|Y}$ after finite sampling do not show algorithmic
dependencies between $S$ and $M$. For continuous variables, this is already obvious from the fact that
the conditional distribution of $X$, given $Y$, 
is only defined for  the support of $q_Y$. 
If the true distribution is a density, the empirical  distribution
contains every $y$-value only once and knowing the support of $q_Y$ thus already implies knowing $q_Y$.

To circumvent this problem, 
we will in the following section consider additive noise  models over a finite probability space. Within  this setting, we derive statements on distributions that are  {\it close to} additive noise models.
Since the finite case has the advantage that
empirical  frequencies converge pointwise to the true probabilities,  this result also implies
statements for the corresponding empirical distribution.

\section{Stronger statements in finite probability spaces}

\label{Dis}

The following theorem is a modification of Theorem~\ref{CAI} for additive noise models over the finite field
$\Z_m$ for some prime number  $m$.

\begin{Thm}[Algorithmic information between $p_Y$ and $p_{X|Y}$ for the discrete model]${}$\\  \label{MIcond}
Let $p_{X,Y}$ be a computable strictly positive distribution on $\Z_m^2$ for some prime number $m$ that admits an additive noise model, i.e.,
there is a function
$f:\Z_m \rightarrow \Z_m$  such that 
$E:=Y-f(X)$  and $X$ are statistically independent.
Here, subtraction is  understood with respect to $\Z_m$.
Then, if $f$ is non-constant, we have 
\begin{equation}\label{InMIexact}
I(p_Y:p_{X|Y}) \stackrel{+}{\geq} K(p_Y)-2 \log m \,.
\end{equation}
\end{Thm}

\noindent
Proof: The idea is, again, to derive an equation that shows that $p_Y$ is essentially determined by 
$p_{X|Y}$ up to some small amount  of additional information.
We have 
\[
\log p(x,y)=\log p_X(x) +\log p_E(y-f(x))\,.
\]
 Defining $\delta:=f(x_0+1)-f(x_0)$, for some $x_0$ for  which  $\delta\neq 0$, we introduce
\begin{equation}\label{kdef}
k_{(x|y)} = \log p(x-1|y)- \log  p(x-1|y-\delta) + \log p(x|y) 
-\log  p(x|y+\delta)\,,
\end{equation}
which
yields the equation
\begin{equation}\label{eqDGLdisc}
\log  p(y+\delta) -  \log p(y) = k_{(x_0|y)}+ \log p(y)- \log p(y-\delta). 
\end{equation}
We interpret eq.~(\ref{eqDGLdisc}) as a  discrete version of eq.~(\ref{DGL}) because 
it relates differences between the values $\log p(y)$ at different points $y$ to the quantity $k_{(x|y)}$, which 
is a property of the conditional $p_{X|Y}$ alone. 
Eq.~(\ref{eqDGLdisc}) implies for arbitrary $y_0$
\[
\log p(y_0+(j+1)\delta) -\log p(y_0+j\delta) =\log p(y_0+j\delta) -\log p(y_0+(j-1)\delta)+ k_{(x_0|y+j\delta)} \,,
\] 
for all $j=1,\dots,m$.
Writing $\log p_Y$  for the vector with coefficients $\log p(y_0+(j+1)\delta)$ 
and  $k$ for the vector with coefficients 
 $k_{(x_0|y+j\delta)}$ for $j=0,\dots,m-1$, we rewrite eq.~(\ref{eqDGLdisc}) as
\[
(S-I)^2 \log p_Y =k\,,
\]
where $S$ denotes the cyclic shift in dimension $m$.  Using the 
fact that $(S-I)$ is invertible on the space of vectors with zero sum of coefficients,
we 
thus obtain
\begin{equation}\label{2Diff}
\log p_Y= (S-I)^{-2} k +\alpha\, {\bf e}\,, 
\end{equation}
where $\alpha$ is given by normalization and ${\bf e}$ is the vector  with only ones as entries.
This shows that $x_0$, $\delta$, and $p_{X|Y}$ determine $p_Y$. Denoting $i:=(x_0,\delta)$  we can summarize the above into
$K(p_Y|p_{X|Y},i) \peq 0$. 
This implies
\[
K(p_Y|p_{X|Y})\pleq K(i)\,,
\]
because 
\begin{eqnarray*}
K(p_Y|p_{X|Y}) - K(p_Y|p_{X|Y},i)
&\peq& K(p_Y|p_{X|Y}) -K(p_Y|p_{X|Y},K(i|p_{X|Y}),i)\\
 &\Ceq &  I(p_Y:i|p_{X|Y}) \pleq K(i)\,,
\end{eqnarray*}
where the second equality  is due  to the definition 
of conditional algorithmic mutual information (\ref{cMI}).
$\Box$

\vspace{0.3cm}
\noindent
We  want to derive a similar lower bound for the case where $p_{XY}$ {\it almost} admits an additive 
noise model. To  this end, we first 
define a precision dependent Kolmogorov complexity  of a probability distribution:

\begin{Def}[Precision dependent algorithmic information]${}$\\
Let $p$ be a density on 
 finite probability space. Let $r$ be a computable probability density and $K(r)$ be the length of the
shortest program on a  universal Turing machine that computes $r(x)$ from $x$.
Then
\[
K_\epsilon (p):=\min_{r \hbox{ with } D(p||r)< \epsilon} K(r|\epsilon)\,,
\]   
where $D(\cdot ||\cdot)$  denotes the relative entropy distance.
Similarly, we define the conditional complexity
$K_\epsilon (p| i)$ given  some prior information $i$. 
\end{Def}

If $q$ is an arbitrary approximation of a distribution $p$ in the sense that  $|\log p(x) - \log q(x)| \leq \epsilon$ holds for all $x$, then $D(p||q)\leq \epsilon$ and thus  the precision dependent algorithmic information can be  bounded from above by the complexity of the approximation: $K_\epsilon(p) \leq K(q)$.
For computable $p$, we obviously have
\[
\lim_{\epsilon\to 0} K_\epsilon (p)=K(p)\,,
\]
but for uncomputable $p$, the complexity tends to infinity. 
The following lemma shows the empirical content  of precision-dependent complexity:

\begin{Lem}[precision-dependent complexity of empirical distributions]${}$\\ \label{emp}
Let $p$ be a positive definite distribution on a finite probability space and $q^{(n)}$ be  the empirical 
distribution after $n$-fold sampling from $p$.
Then 
\[
\lim_{n\to \infty} K_\epsilon (q^{(n)}) = K_\epsilon (p)\,,
\]
with  probability one.
\end{Lem}

\noindent
Proof: Let $r$ be a distribution for which $K_\epsilon(p)=K(r)$ and $D(p||r)<\epsilon$.
due  to $D(q^{(n)}||r)\rightarrow D(p||r)$ with probability  one and because of 
the continuity of relative entropy 
for positive definite distributions   
we also have 
$D(q^{(n)}||r)<\epsilon$  for all  sufficiently large $n$.
Hence $K_\epsilon(q^{(n)})\leq K_\epsilon(p)$.

To prove that $K_\epsilon(q^{(n)})\geq K_\epsilon(p)$, let $r^{(n)}$ be a sequence of distributions  such that
$K_\epsilon(q^{(n)})=K(r^{(n)})$ and $D(q^{(n)}||r^{(n)})<\epsilon$.
Hence,  $D(p||r^{(n)})<\epsilon$ for sufficiently  large $n$ which completes the proof.$\Box$

\vspace{0.3cm}
\noindent
The following lemma will later be used to derive a lower  bound on $I(p_Y:p_{X|Y})$ in terms
of $K_\epsilon(p_Y)$:

\begin{Lem}[mutual information and approximative descriptions]${}$\\ \label{ApMu}
Let $p$ be a computable distribution on a finite probability space, $z$ an arbitrary string and $\epsilon >0$ 
computable.  Let $q$ be a distribution that is $\epsilon$-close to $p$, i.e., 
\begin{equation}\label{rApp}
D(p||q) <  \epsilon\,.
\end{equation}
If $q$ can be derived from $z$ and from $p$  in the sense that 
\begin{equation}\label{rDer}
K(q|p,i_p)\stackrel{+}{=}K(q|z,i_z)\stackrel{+}{=}  0\,,
\end{equation}
for additional strings $i_p$ and $i_z$, then
\[
I(p:z) \stackrel{+}{\geq}  K_\epsilon(p) - K(i_p)-K(i_z).
\]
\end{Lem}

\noindent 
Proof: Using the definition of conditional mutual information (\ref{cMI}) we get
\begin{eqnarray*}
I(q:i_p|p) &\peq& K(q|p) - K(q|i_p,K(i_p|p),p) \peq K(q|p),
\end{eqnarray*}
because Eq.~(\ref{rDer}) implies $K(q|i_p,K(i_p|p),p)\Ceq 0$. On the other hand $I(q:i_p|p) \pleq K(i_p)$ and therefore $$K(q|p) \pleq K(i_p).$$ 
In the same way,
eq.~(\ref{rDer}) implies $K(q|z) \pleq K(i_z)$. A data processing inequality (Corrolary II.8 in \cite{GacsTromp}) then implies
\[
I(p:z)\stackrel{+}{\geq} K(q) - K(i_p)-K(i_z)\,.
\] 
We conclude with $K_\epsilon (p)\stackrel{+}{\leq} K(q)$ due to ineq.~(\ref{rApp}).
$\Box$

\vspace{0.3cm}
\noindent
We  will moreover need  the following Lemma:
\begin{Lem}[bound  on the differences of logarithms]${}$\label{lemLogBound}\\
Given a vector $v\in  \R^m$, we define a probability distribution by
\[
p_j:=\frac{1}{z_v} e^{-v_j} \,,
\]
where $z_v$ is  the partition function.  Let $\tilde{p}$ be defined by $\tilde{v}$ in the same way.
Then 
\[
|\log p_j -\log \tilde{p}_j |\leq 2\|v-\tilde{v}\|_\infty\,.
\]
\end{Lem}

\noindent
Proof: Due to
\[
\log p_j -\log \tilde{p}_j = v_j -\tilde{v}_j - \log z_v +\log z_{\tilde{v}}
\]
we only have to show
\[
 |\log z_v -\log z_{\tilde{v}}| \leq \|v-\tilde{v}\|_\infty\,.
\]
To this end, we define
\[
\log z(\epsilon):= \log z_{v+\epsilon (\tilde{v}-v)}\,.
\]
Using the mean value theorem  we have for an appropriate value $\eta \in (0,1)$ 
\begin{eqnarray*}
\log z_{\tilde{v}}-\log z_v &=&\log z(1)-\log z(0)\\
                  &=&(\log z)'(\eta)\\
                  &=& \sum_j (v_j -\tilde{v}_j) \frac{1}{z(\eta)} e^{-v_j+\eta (v_j-\tilde{v}_j)}\,.
\end{eqnarray*}
The  last expression is the expected value of $v_j-\tilde{v}_j$ with respect to the probability distribution
corresponding to $v+\eta (\tilde{v}-v)$, which cannot be greater than $\|v-\tilde{v}\|_\infty$.
$\Box$

\vspace{0.3cm}
\noindent
We now have introduced the technical requirements to formulate  a theorem for
approximate additive noise models:

\begin{Thm}[approximate additive noise model]${}$\\ \label{Apan} 
Let $p_{X,Y}$ be as in Theorem~\ref{MIcond}, but only admitting 
an approximative additive noise model in the sense that
\begin{equation}\label{mu}
I(X:E) \leq \frac{\beta}{2} \left(\frac{\epsilon \beta}{4 m^3}\right)^2\,,
\end{equation}
where $\beta$ is a lower bound on $p(x,y)$.
Here, subtraction is  understood with respect to $\Z_m$.
Then, if $f$ is non-constant, we have 
\begin{equation}\label{InMI}
I(p_Y:p_{X|Y}) \stackrel{+}{\geq} K_{\epsilon} (p_Y)- 2\log m  - m -2K(\epsilon)\,.
\end{equation}
\end{Thm}
\vspace{0.2cm}
\noindent
Proof: The idea is to define a distribution  $\tilde{p}_{X,Y}$ that is close to $p_{X,Y}$ and admits an {\it exact} additive noise model:
Define a joint distribution on $X$ and  $E$ by the product
\[
\tilde{p}_{X,E}:=p_X \otimes p_E\,.
\]
By variable transformation, $\tilde{p}_{X,E}$ defines a distribution $\tilde{p}_{X,Y}$ that admits an additive noise model from $X$ to $Y$.
Eq.~(\ref{eqDGLdisc})  now holds  for
$\tilde{p}_{X|Y}$ and  $\tilde{p}_Y$ with $\tilde{k}_{(x_0|y)}$ instead of $k_{(x_0|y)}$, which is defined similar to  eq.~(\ref{kdef}). Denote the corresponding vector by $\tilde{k}=(\tilde{k}_{(x_0|y)})_y$.
In analogy to eq.~(\ref{2Diff}) and the proof of Theorem~\ref{MIcond}, we now have
\[
\log \tilde{p}_Y= (S-I)^{-2} \tilde{k} +\tilde{\alpha} {\bf e}\,,
\]
where $\tilde{\alpha}$ is the appropriate normalization constant and ${\bf e}$ the all-one vector.
To show that $p_{X|Y}$ allows an approximative description of $p_Y$ we have  to replace
$\tilde{k}$ and  $\tilde{p}_Y$ with $k$ and $p_Y$, respectively. 
We define 
\[
\log r_Y:=(S-I)^{-2}k +\alpha {\bf e}\,,
\]
and using Lemma \ref{lemLogBound} we  obtain
\begin{eqnarray}
\|\log p_Y - \log r_Y\|_\infty &\leq & \|\log p_Y - \log \tilde{p}_Y\|_\infty + \|\log \tilde{p}_Y - \log r_Y\|_\infty \nonumber \\
&\leq & \|\log p_Y - \log \tilde{p}_Y\|_\infty + 2 \|(S-I)^{-2}(k-\tilde{k})\|_\infty. \label{b1}
\end{eqnarray}
The modulus of the eigenvalues of 
$(S-I)^{-1}$ on this subspace are  all smaller than $m/4$ (for $m\geq 2$) since they read
\[
\frac{1}{e^{2\pi\, i/m}-1}, \frac{1}{e^{2\pi  i\, 2/m}-1}      ,\ldots,\frac{1}{e^{2\pi i\, (m-1)/m}-1}\,.
\]
We thus have
\[
\|(S-I)^{-2} (\tilde{k} - k)\|_2 \leq \frac{m^2}{16} \|\tilde{k}-k\|_2\leq \frac{m^3}{16} \|\tilde{k}-k\|_\infty \,,
\]
where the last  inequality used $\|\cdot \|_2\leq \sqrt{m} \|\cdot \|_\infty$.
Together with $\|\cdot \|_\infty \leq \|\cdot \|_2$, ineq.~(\ref{b1}) then yields
\begin{equation}\label{p_tilde_p}
\|\log p_Y - \log r_Y\|_\infty \leq \|\log p_Y - \log \tilde{p}_Y\|_\infty + \frac{m^3}{8}\|\tilde{k} - k \|_\infty.
\end{equation}
Now we derive an upper bound on the two summands of the rhs.~using our assumption on the limited statistical mutual information between  $X$ and $E$. To this end, we observe that
\begin{equation}\label{monot}
D(p_{X,Y}||\tilde{p}_{X,Y})= D(p_{X,E}||\tilde{p}_{X,E})=I(X:E)\,,
\end{equation}
where the first equality is due  to the invariance of relative entropy under variable transformation and the second uses a well-known reformulation  of mutual information \cite{Cover}.
Moreover, we have 
\[
D(p_{X|Y}||\tilde{p}_{X|Y})=\sum_y D(p_{X|y}||\tilde{p}_{X|y}) p(y) \leq \frac{\beta}{2} \left(\frac{\epsilon\beta}{ 4 m^3}\right)^2 \,,
\]
where $p_{X|y}$ denotes  the conditional distribution for one specific value $y$  of $Y$.
Using the lower bound on $p(y)$ we obtain
\[
 D(p_{X|y}||\tilde{p}_{X|y}) \leq  \frac{1}{2}\left(\frac{\epsilon\beta}{ 4 m^3}\right)^2     \quad \forall y\,.
\]
Due to the well-known  relation $D(p||q)\geq (2\ln 2)^{-1}\|p-q\|^2_1$ between relative entropy and $\ell_1$-distance  
for two distributions \cite{Cover}, we obtain
\[
|p(x|y)-\tilde{p}(x|y)| \leq \frac{\epsilon \beta}{ 4 m^3} \,.
\]
This implies
\begin{equation}\label{tildeErrorCon}
|\log p(x|y)-\log \tilde{p}(x|y)| \leq \frac{\epsilon}{ 4 m^3}    \,,
\end{equation}
by applying  the mean value theorem to the function $a \mapsto \log  a$.
From the definition of  $\tilde{k}_{(x|y)}$ and $k_{(x|y)}$ in
eq.~(\ref{kdef}) we conclude
\begin{equation}\label{kerror}
\|\tilde{k}-k\|_\infty  \leq \frac{\epsilon}{ m^3}\,.
\end{equation}
On the other hand, ~(\ref{monot}) implies 
\[
D(p_Y||\tilde{p}_Y)\leq \frac{\beta}{2} \left(\frac{\epsilon\beta}{ 4 m^3}\right)^2 \leq \frac{1}{2}\left(\frac{\epsilon\beta}{ 4 m^3}\right)^2  \,,
\]
and hence
\begin{equation}\label{tildeError}
\|\log p(y)-\log \tilde{p}(y)\|_\infty\leq \frac{\epsilon \beta}{ 4 m^3} < \frac{\epsilon}{8}.
\end{equation}
Using  ineqs.~(\ref{kerror}) and ~(\ref{tildeError}), ineq.~(\ref{p_tilde_p}) yields for all $y$
\begin{equation}\label{final_bound}
|\log p(y)- \log r(y)| < \frac{\epsilon}{4}\,.
\end{equation}
Let $\log q_p(y)$ be given by discretizing all values $\log p(y)$ up to an accuracy of $\epsilon/4$. Then 
$$
K(q_p|p_Y,\epsilon)\Ceq 0.
$$
On the other hand, let $\log q_r(y)$ be given by discretizing all values $\log r(y)$ up to an accuracy of $\epsilon/4$. Then $K(q_r|r,\epsilon)\Ceq 0$ and thus 
\[
K(q_r|p_{X|Y},\delta,x_0,\epsilon)\Ceq  0\,.
\]
Due to (\ref{final_bound}), both discretizations coincide up to one bit for each value $y$, say $b_m(y)$. To illustrate this, consider the binary strings $0.111\ldots$ and $1.000\ldots$ which can be arbitrarily close despite their truncation being different. We conclude that 
$$
K(q_p|p_{X|Y},\delta,x_0,\epsilon,b_m)\Ceq  0.
$$
Let $q$ be the distribution generated by $\log q_p$ through normalization
$$
\log q(y) := \log q_p - \log \sum_y q_p(y).
$$
Due to the upper bound (\ref{final_bound}), Lemma \ref{lemLogBound} gives
$$
D(p||q) \leq 2 \|\log p(y)-\log q_p(y)\|_\infty < \epsilon.
$$
The theorem now follows from  Lemma~\ref{ApMu} applied to
$
z=p_{X|Y},\; i_z=(\delta,x_0,\epsilon,b_m)\;, p=p_Y\; \text{ and } i_p = \epsilon\,.
$
$\Box$

\vspace{0.3cm}
\noindent 
The  complexity of $p_Y$ in the bound (\ref{InMI}) will typically exceed the terms with $m$ 
because  we will need several bits for every bin to describe the corresponding probability (this will be discussed
in  Section~\ref{EstK} in more detail).
Moreover, $K(\epsilon)$ can be quite low, in particular if  we choose $\epsilon =2^{-k}$ for some  $k$.
Therefore, the mutual information between $p_Y$ and $p_{X|Y}$ is almost as large as the complexity
of $p_Y$. This shows that  the amount of adjustments required to mimic an additive noise model in the wrong 
direction depends essentially on the complexity  of $p_Y$. 
In the following section we consider the complexity in the case in which $p_Y$ is typical with respect to some known parametric family of distributions.

\section{Kolmogorov complexity of distributions from a parametric family}

\label{EstK}

The problem with applying Theorems~\ref{MIcond} and \ref{Apan} to real data is  that  
the 
term $K_\epsilon (p_Y)$ cannot be known due to the  uncomputability of Kolmogorov complexity in  general.
Fortunately, we can prove statements about the increase  of the complexity 
for decreasing  $\epsilon$ 
for typical elements of a {\it family} of distributions.
This is shown
by the following lemma:

\begin{Lem}[typical distributions in parametric families]${}$\\ \label{typ}
Let $p_\theta$ be a parametric family of distributions over some finite probability space and  $\theta$
run over a $d$-dimensional manifold $\Lambda\subset \R^d$.
Moreover, let  $p_\theta$  be computable in the following sense:
there exists a program that computes $p_\theta(y)$ for any computable input  $\theta$.
If the
Fisher  information matrix has 
full rank for all $\theta \in  \Lambda$, the complexity of a typical distribution 
$p_\theta$  grows logarithmically with  decreasing $\epsilon$, i.e. for sufficiently small $\epsilon$
\[
K_\epsilon (p_{\theta}) \stackrel{+}{=} - \frac{d}{2} \log  \epsilon\,.
\]
\end{Lem}

\noindent
Proof: Let $F_\theta$ denote the Fisher information matrix of the parametric family  
and 
$\theta_1,$ $\theta_2,$ $\dots,$ $\theta_{N(k)}  \in \Lambda$ be the parameter vectors of all computable distributions $p_\theta$
that can be described with complexity $K(p_\theta)\leq k$. 

For every $\theta_j$ we have \cite{Cover}
\begin{equation}\label{REF}
D(p_\theta  || p_{\theta_j})= (\theta-\theta_j)^T F_{\theta_j} (\theta -\theta_j) +O(\|\theta -\theta_j\|^3)\,.
\end{equation}
For sufficiently small $\epsilon$, the set of all $\theta$ with $D(p_\theta||p_{\theta_j})\leq \epsilon$ is
thus contained in the ellipsoid 
\[
(\theta-\theta_j)^T F_{\theta_j} (\theta -\theta_j) \leq 2   \epsilon\,.
\]
The volume $V_j$ of such an ellipsoid with respect to the Lebesque measure is given by
\[
V_j=(\det F_{\theta_j})^{-1/2}   \frac{\pi^{d/2}}{\Gamma (d/2+1)} (2 \epsilon)^{d/2}\,.
\] 
This can be seen by transforming the ellipsoid into a sphere of radius $\sqrt{2\epsilon}$  
via the linear map $(F_{\theta_j})^{-1/2}$. 

Now we check how the minimum number of disjoint ellipsoids must increase with $\epsilon$ 
to cover at least a constant fraction of the parameter space $\Lambda$. 
Otherwise, if the total volume 
tends to zero  it  gets more and more unlikely  that it contains a randomly chosen $\theta\in \Lambda$. 
We need  to increase $N(k)$ proportional 
to $1/(2\epsilon)^{d/2}$ and $k$ must increase with $-\frac{d}{2} \log \epsilon$ due to $N(n)\leq 2^k$. 
Hence we need asymptotically at least $-(d/2) \log_2 \epsilon $ bits.

To see that this is also sufficient, we consider a cube $[0,\lambda]^d\supseteq \Lambda$ that we divide  
 into $N$ equally sized cubes  of side length $\Delta$
with middle points $\theta_1,\dots,\theta_N$
such that 
\[
(\theta  -\theta_j)^T  F_\theta (\theta-\theta_j) \leq \epsilon/2
\]
for any point $\theta$ in the same cube. By (\ref{REF}), this ensures for all $\theta,\theta_j\in \Lambda$ 
and sufficiently small $\epsilon$
that $D(p_\theta|p_{\theta_j})\leq \epsilon$.  
If $\mu$ is an upper bound for  all eigenvalues of all $F_\theta$ it is sufficient to guarantee
\[
\|\theta-\theta_j\|^2  \leq \frac{\epsilon}{2\mu}\,. 
\]    
This can be  achieved by choosing
\[
\Delta \leq \sqrt{\frac{\epsilon}{2\mu d}}\,.
\]
Hence it is sufficient to choose the smallest $N$ that satisfies
\[
N \geq  \left(\frac{\epsilon}{2\mu d}\right)^{d/2}\,,
\]
and whose $d$th root is integer.
The grid and  thus every  vector $\theta_j$ can be computed from $\lambda$ and $j$
and $p_{\theta_j}$ can be computed from $\theta_j$ by assumption. Hence, 
\[
K(p_{\theta_j})\stackrel{+}{\leq }\log_2 N \stackrel{+}{=}- \frac{d}{2} \log \epsilon\,.
\]
$\Box$

\vspace{0.2cm}
\noindent
We will now apply Lemma~\ref{typ} to the family of all distributions $p_Y$ for which $p(y)$ is bounded from below 
by
some $\beta>0$. It is canonically parameterized by the first $m-1$ probabilities if there 
are $m$ possible $y$-values. Then we obtain:

\begin{Cor}[algorithmic mutual information for typical distributions]${}$\\
Let $p_{X,Y}$ be as in Theorem~\ref{Apan}. Further assume that $p_Y$ is typical in the family of distributions on $m$-values whose probabilities are bounded from below by some
$\beta>0$. If $I(X:E)$ satisfies the bound (\ref{mu}) with $\epsilon=2^{-N}$ for sufficiently large $N$, then 
$$
I(p_Y:p_{X|Y}) \stackrel{+}{\geq} \frac{m-5}{2}\log N - 2\log m  -m\,.
$$
\end{Cor}
\noindent
Proof: One can check that the Fisher information matrix 
has full rank.
Then the proof of the preceding lemma shows for sufficiently small $\epsilon$
\[
K_\epsilon(p_Y) \pgeq -\frac{m-1}{2} \log \epsilon\,.
\] 
Plugging this into the lower bound of Theorem~\ref{Apan} together with $\epsilon=2^{-N}$ concludes the proof.
$\Box$

\vspace{0.3cm}
\noindent
Hence, for typical $p_Y$, the lower bound is positive if $m$ and $n$ are large enough.  This  asymptotic statement  still holds true if $p_Y$ looks on a coarse-grained scale like some simple distribution $q_Y$, i.e., a Gaussian, but 
shows irregular deviations from   $q_Y$ if the probabilities are described more accurately.

To give an impression on the 
amount of information 
between
 $p_{Y|X}$ and  $p_Y$ 
that can be inferred after $n$-fold sampling, 
we  recall that  
the mutual information between $E$ and $X$ can be estimated up to an accuracy  of $O(1/n)$ \cite{Paninski}.
The lowest upper bound on $\epsilon$ in ineq.~(\ref{mu}) that can be  guaranteed by the observations 
thus is proportional to $1/\sqrt{n}$. Hence, for constant $m$, the best lower bound on  the amount   of
algorithmic information shared  by $p_Y$ and $p_{X|Y}$ 
increases logarithmically in $n$ as long as the sample is not sufficient to reject independence
between $Y-f(X)$ and $X$.

\section{Applying the results to empiricial distributions}\label{secEmp}

In applying Theorems~\ref{MIcond} and \ref{Apan} to realistic situations, we still have the problem that 
we have  no reason to believe that the true distribution is computable.
On the other  hand, applying the argument to the empirical distribution 
(which is, for large sampling close to an additive noise model) is still problematic
because algorithmic dependencies
between the empirical distribution $q_Y$ and the empirical conditional $q_{X|Y}$ 
do not prove algorithmic dependencies between  the true distributions $p_Y$ and $p_{X|Y}$.
One reason is that every conditional probability 
$q_{Y|X}(y|x)$ can always be written as  a fraction with denominator $q_X(x)n$, which already is an algorithmic dependence.

We now describe how to 
use Postulate~\ref{IC} if only a finite list of 
 $(x,y)$-pairs  is observed and the underlying distribution is not known. 
Given samples
$\mathcal{S}_n=\big[(x_1,y_1),\dots,(x_n,y_n)\big]$,
we can generate a non-empty subsample $\mathcal{S}_{\ell(n)} = \big[(x_1,y_1),\dots,(x_{\ell(n)},y_{\ell(n)})\big]$ with high probability such that every $x$-value occurs exactly $\ell(n)/m$-times. The samples $S_{\ell(n)}$ can then be used for the 
estimation of $p_{Y|X}$. 
Hereby, $\ell(n)$ is chosen independently of the samples in a way that for $n\rightarrow \infty$ we have $\ell(n)\rightarrow 
\infty$ and the probability of obtaining $\mathcal{S}_{\ell(n)}$ from $\mathcal{S}_n$ converges to one.

Now by construction, if $M$ contains no information about $S$, the empirical distribution 
\[
q^{(\ell(n))}_{Y|X}
\]
 of the  {\it subsample} must not contain any information about the empirical distribution 
\[
q^{(n)}_X
\]
 of $x$-values in the entire sample, i.e.,
\begin{equation}\label{ICformal}
M_{X\rightarrow Y}:=I(q^{(n)}_X:q_{Y|X}^{(\ell(n))})\approx 0\,.
\end{equation}
In the spirit of \cite{LemeireD}, we postulate that the violation of eq.~(\ref{ICformal}) 
indicates that the causal hypothesis $X\rightarrow Y$  is wrong
or the mechanisms generating $x$-values and  the mechanisms generating $y$-values from $x$-values
have not been generated independently. For a discussion of this case see \cite{LemeireNIPS}.
Using this terminology, our goal is to derive a lower bound on  $M_{Y\rightarrow X}$ for
the case where $p_{X,Y}$ admits an additive noise model from $X$ to $Y$. 
We can apply Theorem~\ref{Apan} to a distribution that is defined by the empirical results via
\[
p'(x,y):=q^{(n)}(y) q^{(\ell(n))}(x|y)\,,
\]
which is necessarily  computable because it only contains rationale values.
 
We have already argued that the causal hypothesis $Y\rightarrow X$ would only be acceptable if
\[ 
I(q^{(n)}(y) :q^{(\ell(n))}(x|y)) \approx 0\,.
\]
 If the true distribution $p$ almost admits an
additive noise model from $X$ to $Y$ in the sense of ineq.~(\ref{mu}), the same inequality will also be satisfied by
$p'$ if $n$ is sufficiently high and thus  
\[
I(q^{(n)}_Y :q^{(\ell(n))}_{X|Y})\gg 0
\]
provided that $K_\epsilon(q^{(n)}_Y)$, which coincides with  $K_\epsilon (p_Y)$ due  to Lemma~\ref{emp} for large $n$, is high.

\section{Approximating  continuous variables  with discrete ones}

\label{Con}

Causal inference via
additive noise models has been described and tested for continuous variables \cite{Hoyer}.
We have discussed the discrete case mainly for technical reasons because we were able to prove
statements for distributions that  are only close  to additive noise models.
Our results  can easily be applied to the  continuous case by discretization with increasing number of bins. 
As already mentioned, the discretized version of the empirical distribution becomes computable, which 
circumvents the problem that the true distribution may be uncomputable.

Before we discuss the discretization in detail, we emphasize that there is a problem with applying
Postulate~\ref{IC} to the conditionals obtained after discretizing the variables:
if we define a discrete variables $X^{(m)}$ and $Y^{(m)}$  by putting $X$ and $Y$ into $m$ bins each, the
discretized conditional $p_{Y^{(m)}|X^{(m)}}$ does not only depend  on $p_{Y|X}$. Instead,  it also contains
information about the distribution of $X$.  For this reason,
algorithmic dependencies between $p_{Y^{(m)}|X^{(m)}}$ and  $p_{X^{(m)}}$ only disprove the causal
hypothesis $X\rightarrow Y$ if the binning is fine enough to guarantee that the discrete value $x^{(m)}$ is
sufficient to  determine  the conditional probability for $y^{(m)}$, i.e., the relevance of the exact value 
$x$ is negligible if the discrete value is given.
It is therefore essential that the argument below  refers to  the asymptotic case
of infinitely small  binning.

To approximate a continuous density $p(x,y)$  on $\R^2$ by $\Z_m^2$ with increasing  $m:=2k+1$ we consider 
the square 
\[
Q_m:=\left[-\frac{1}{2}\sqrt{m}, \frac{1}{2}\sqrt{m}\right]^2
\]
for all odd $m$ and replace $p(x,y)$ with
$p(x,y|Q)$. We discretize $Q$ into $m\times m$ bins of equal size, which defines a probability distribution over
$\Z_m$-valued variables $X_m$ and $Y_m$,  respectively. 
We define the function $f_m:\Z_m\rightarrow \Z_m$ by putting the values $f(\Delta (j-1/2))$ with $j=-k,\dots,k$
to  the corresponding bin.   

Moreover, appropriate smoothness asumptions on $p(x,y)$ can guarantee
that the  mutual information between $Y_m-f_m(X_m)$ and $X_m$
converges to $I(X:(Y-f(X)))$ for $m \to \infty$. 
It is known \cite{Paninski} that there are estimators for mutual information that converge if
the binning $m$ is increased proportionally to $\sqrt{n}$ for sample size  $n\to\infty$. 
If $p(x,y)$ admits an additive noise model, i.e.,  $I(X:(Y-f(X))  =0$, then  
$I(X_m:(Y_m-f(X_m))\rightarrow 0$. Hence, the discrete distributions on $X_m$ and  $Y_m$  
get arbitrarily close to discrete additive noise models. Applying Theorem~\ref{MIcond} to these discrete distributions then yields algorithmic dependence between
the discretized marginal and the discretized conditional.

\section{Conclusions}

We  have discussed a causal inference method that prefers the causal hypothesis 
$X\rightarrow Y$ to  $Y\rightarrow X$ whenever
the joint distribution  $p_{X,Y}$ admits an additive noise model from $X$ to $Y$ and not vice versa. 
It seems that this way of reasoning assumes that all causal mechanisms in nature can be described 
by additive noise models  (which is certainly not the case). Here we argue that the method is nevertheless
justified because it is unlikely that a causal mechanism that is not of the form of an additive noise model
generates a distribution  that looks like an additive noise model in the {\it wrong} direction.
This is because such a coincidence would require mutual adjustments between $P({\rm cause})$ and $P({\rm effect}|{\rm cause})$.
To measure the amount of tuning needed for this situation we have derived a lower bound
on the algorithmic information shared by $P({\rm cause})$ and $P({\rm effect}|{\rm  cause})$. If we assume that
``nature chooses'' $P({\rm cause})$ and $P({\rm effect}|{\rm cause})$ independently, a significant amount of algorithmic
information is not acceptable. Our justification of additive-noise-model based causal discovery thus
is an  application of two recent  proposals for using algorithmic information theory in causal inference:
\cite{LemeireD} postulated that the shortest description of $P({\rm cause},{\rm effect})$ is given  by
separate descriptions of $P({\rm cause})$ and $P({\rm effect}|{\rm cause})$, which would be violated then.
\cite{Algorithmic} argued that algorithmic dependencies between any two objects require a causal explanation.
They consider the two mechanisms that determine 
$P({\rm cause})$ and $P({\rm effect}|{\rm cause})$, respectively, as two objects and conclude that  the absence
of causal links on the level of the two mechanisms imply their algorithmic independence, in agreement 
with \cite{LemeireD}.


\begin{thebibliography}{10}

\bibitem{Chaitin}
G.~Chaitin.
\newblock On the length of programs for computing finite binary sequences.
\newblock {\em J. Assoc. Comput. Mach.}, 13:547--569, 1966.

\bibitem{ChaitinF}
G.~Chaitin.
\newblock A theory of program size formally identical to information theory.
\newblock {\em J. Assoc. Comput. Mach.}, 22:329--340, 1975.

\bibitem{Cover}
T.~Cover and J.~Thomas.
\newblock {\em Elements of Information Theory}.
\newblock Wileys Series in Telecommunications, New York, 1991.

\bibitem{GacsTromp}
P.~Gacs, J.~Tromp, and P.~Vit\'{a}nyi.
\newblock Algorithmic statistics.
\newblock {\em IEEE Trans. Inf. Theory}, 47(6):2443--2463, 2001.

\bibitem{Hoyer}
P.~Hoyer, D.~Janzing, J.~Mooij, J.~Peters, and B~Sch\"olkopf.
\newblock Nonlinear causal discovery with additive noise models.
\newblock In D.~Koller, D.~Schuurmans, Y.~Bengio, and L.~Bottou, editors, {\em
  Proceedings of the conference Neural Information Processing Systems (NIPS)
  2008}, Vancouver, Canada, 2009. MIT Press.
\newblock \url{http://books.nips.cc/papers/files/nips21/NIPS2008_0266.pdf}.

\bibitem{Algorithmic}
D.~Janzing and B.~Sch\"olkopf.
\newblock {Causal inference using the algorithmic Markov condition}.
\newblock {\em {{\tt http://arxiv.org/abs/0804.3678}}}, 2008.

\bibitem{KolmoOr}
A.~Kolmogorov.
\newblock Three approaches to the quantitative definition of information.
\newblock {\em Problems Inform. Transmission}, 1(1):1--7, 1965.

\bibitem{Lauritzen}
S.~Lauritzen.
\newblock {\em Graphical Models}.
\newblock Clarendon Press, Oxford, New York, {Oxford Statistical Science
  Series} edition, 1996.

\bibitem{LemeireD}
J.~Lemeire and E.~Dirkx.
\newblock Causal models as minimal descriptions of multivariate systems.
\newblock {\em {{\tt http://parallel.vub.ac.be/$\sim$jan/}}}, 2007.

\bibitem{LemeireNIPS}
J.~Lemeire and K.~Steenhaut.
\newblock Inference of graphical causal models: Representing the meaningful
  information of probability distribution.
\newblock {\em \em{To appear in} Proceedings of the NIPS 2008 workshop
  ``Causality - Objectives and Assessment'', edited by I.~Guyon, D.~Janzing,
  B.~Sch\"olkopf.}, 2009.

\bibitem{Vitanyi97}
M.~Li and P.~Vit\'{a}nyi.
\newblock {\em An Introduction to Kolmogorov Complexity and its Applications}.
\newblock Springer, New York, 1997.

\bibitem{Paninski}
L.~Paninski.
\newblock Estimation of entropy and mutual information.
\newblock {\em Neural Computation}, 15:1191--1254, 2003.

\bibitem{Pearl:00}
J.~Pearl.
\newblock {\em Causality}.
\newblock Cambridge University Press, 2000.

\bibitem{Reichenbach}
H.~Reichenbach.
\newblock {\em The direction of time}.
\newblock Dover, 1999.

\bibitem{Solomonoff}
R.~Solomonoff.
\newblock A preliminary report on a general theory of inductive inference.
\newblock {\em Technical report V-131}, Report ZTB-138 Zator Co., 1960.

\bibitem{Spirtes1993}
P.~Spirtes, C.~Glymour, and R.~Scheines.
\newblock {\em Causation, prediction, and search (Lecture notes in
  statistics)}.
\newblock Springer-Verlag, New York, NY, 1993.

\bibitem{Zhang_UAI}
K.~Zhang and A.~Hyv\"arinen.
\newblock On the identifiability of the post-nonlinear causal model.
\newblock In {\em Proceedings of the 25th Conference on Uncertainty in
  Artificial Intelligence}, Montreal, Canada, 2009.

\end{thebibliography}

\end{document}